\documentclass[aps,showpacs,toolkits,nofootinbib,preprint]{revtex4}

\usepackage{latexsym}
\usepackage[hypertex]{hyperref}
\usepackage{amssymb}
\usepackage{mathrsfs}
\usepackage{amsmath}

\begin{document}

\title{ Some Comments on the `Quirks' Scenario   }

\author{Shmuel Nussinov}
\email{nussinov@post.tau.ac.il}

\affiliation{Tel Aviv University, Sackler School Faculty of Sciences, \\ Tel Aviv 69978, and Schmid Science Center
Chapman University,\\Orange California  }

\author{Chen Jacoby}
\email{chj3@post.tau.ac.il}

\affiliation{Tel Aviv University, Sackler School Faculty of Sciences, \\Ramat Aviv, Tel Aviv 69978, Israel}

\begin{abstract}

We discuss various aspects of models with long-lived or stable colored particles.
In particular we focus on an ideal Quirk model with electroweak neutral heavy (O(TeV))
particles which carry ordinary color and another $ SU'(3)$ color with a very low scale
$\Lambda'$. We show that contrary to what one might think, such a model is cosmologically
consistent and evades many ``Pitfalls" even for very low O(10 eV) $\Lambda'$
and without assuming a low reheat temperature.
We also show that the expected production of Quirks by cosmic rays  which are
incorporated in heavy Isotopes in Ocean water is consistent with the highly stringent bounds
on the latter. This evades a real threat to the Quirk model which would have excluded
it regardless of Cosmology.  Finally we briefly comment on possible LHC signatures.

\end{abstract}

%\pacs{ }

\maketitle
\vspace{.25in}

\section{Introduction}
Long-lived massive colored particles have been suggested in a ``Split SUSY"
context\cite{AHSDimopolous}. More recently, massive ``Quirks", ($m_{Q'} > 0.5$ TeV) have been
suggested \cite{KangLuty}.  In the ``purest" form Quirks are electroweak neutral but carry
color and another $SU(N')$ color with a very low scale ($\Lambda'$= 10 MeV - 10 eV) and are
{\it stable}.  Such particles have reasonably large production cross sections at LHC
and the ``long" color$'$ strings between them can dramatically manifest therein.

While Quirks do not resolve any astro-particle dilemma and, in particular,  cannot serve as dark matter,
their phenomenology is extremely interesting. We found \cite{JacobyNussinov} that quirks
{\it are} cosmologically viable even without (as Luty and Kang later did) sub-TeV
post-inflation reheat temperatures.  There are a dozen or so different pitfalls, each of
which could potentially exclude this scenario where the $SU(N')$ degrees of freedom (i.e.,
the Quirks and associated gluons) were in thermal equilibrium with all other fields at
some early epoch. We find, however, the model (for $N'$=3) evades all  of them.

The cosmological or other problems facing  Quirk scenarios fall into several categories:

\subsection {Early Freeze-out}
The early freeze-out of Quirk annihilations at temperatures $T \sim m_{Q'}/30$ leaves
 a relic density of Quirks:
\begin{equation}
    Y(Q')=\frac{n_{Q'}}{s} \simeq \frac{n_{Q'}}{n_\gamma} \sim 10^{-14}.
\end{equation}
If all these relic Quirks survive until $T \sim \Lambda'$, when extensive annihilations
follow after forming  $Q'\bar{Q'}$ bound
states confined by $SU(N')$ strings, then some observable signatures in the microwave background
and nuclear abundances are expected. Such signatures have never been
observed. This can be avoided if at $T \sim \Lambda_{\rm{QCD}}$ ``Hadronic Assisted
Annihilations"(``HAA")  further reduces $Y(Q')$ by $10^{-4}$ down to $10^{-18}$.

  %II. Potential..

\subsection{Potential Difficulties due to the New Gluons and the New Glueballs}
  The $N'^2$-1 gluons associated with the new low scale gauge group can be problematic.
  First for confinement scales $\Lambda' <$ MeV these thermal gluons are additional light
  degrees of freedom which can adversely effect the successful predictions of Nucleosynthesis.
  The second, more serious difficulty is that after $T$ drops below $\Lambda'$
  all these $g$'s bind to form $g'g'$ glueballs of masses $\sim \mathcal{O}(\Lambda')$ with
  number density $\mathcal{O}(T'^3) \sim \mathcal{O}(\Lambda'^3)$. The lowest $g'$-balls are
  stable and one needs to verify that their density decreases enough via
  3 $\rightarrow$ 2 processes so as to avoid over-closure.

   %III. Very Late....

\subsection{Very Late Annihilation at $T' < \Lambda'$ due to $SU(N')$ Strings}
  The residual density of Quirks is strongly restricted by the severe terrestrial limits
  on the abundance of anomalous heavy and/or fractionally charged isotopes obtained
  when Quirks bind to nuclei. We have to verify that $SU(N')$ strings connecting $Q'$ and
  $\bar{Q'}$ after the temperature dropped below $< \Lambda'$ indeed guarantee $Q'-\bar{Q'}$
  annihilation which is efficient enough to meet such limits.

   %IV.  Even if ...

\subsection{Heavy Isotopes Due to Quirk Production by Cosmic Rays}
  Even if all Cosmological constraints are met (or simply evaded as in \cite{KangLuty} by
  postulating a low reheat temperature) the following difficulty must be addressed:
  High energy cosmic rays keep producing $Q'-\bar{Q'}$ pairs, at a larger rate,
  due to the stronger color couplings and the extra $N'$ factor, than other uncolored
  putative stable $X-\bar{X}$ pairs of the same mass.
  These $Q$'s accumulate over billions of years in ocean waters as ultra-heavy isotopes.
  For the viability of {\it any} quirk scenario, we must therefore show that:  The cosmic ray
  produced $Q$'s and $\bar{Q}$'s even when incorporated into heavy isotopes in
  water, annihilate efficiently, evading this possible difficulty as well.

   %IV. Direct....

\subsection{ Direct Signatures of $Q'-\bar{Q'}$ Production in the LHC Detectors}
  Anticipating that this will be the thrust of the (then) future paper by Kang and Luty, we
  have only briefly discussed it in our original paper. However, we find, as we detail below,
  that some of these signatures may be even more dramatic and striking and can be better
  estimated than what was pointed out by Luty and Kang.

   The all-important issue of Quirk production by cosmic rays (raised by W.~Marciano in a
  seminar given by one of us at BNL) has not been addressed in the context of Quirk models
  before.
  This and other remarks regarding difficulties of the cosmological scenario with abundant
  initial Quirks and $g$'s prompted us to address and clarify the above key points.
  In \cite{JacobyNussinov} we discussed also the remote yet fascinating possibility arising
  for rather low $\Lambda$'s that Quirks produced at LHC and stopped in its vicinity can
  allow us to directly study and manipulate long  strings.  We do not elaborate on this here.

   %I. Estimating ...

\subsection{Estimating Relic Density Reduction via HAA (Hadronic Assisted Annihilation)}
  Extensive literature quoted in \cite{Kang:2006yd} shows that reducing the relic abundance
  of the heavy $M \sim$ TeV colored particles by about $4$ orders of magnitude below the
  $Y(Q') \sim 10^{-14}$ expected from early perturbative annihilation allows lengthening
  of their allowed lifetimes all the way to $10^{14}$ sec.

  In our cosmological Quirk scenario, practically all relic $Q$'s which survive the early and
  Hadronic Assisted Annihilations (HAA), annihilate shortly after being confined by color$'$
  strings at $T' \sim \Lambda'$. For confinement scales $\Lambda' \sim$ 10 eV - 10 KeV  this
  happens at times  $t \sim 10^{10} - 10^4$ sec far shorter than the above-mentioned $10^{14} $
  sec allowed region for a $10^{-4}$ HAA reduction.  Therefore even smaller reductions via
  HAA of $Y(Q')$ by $\sim 10^{-3}-10^{-2}$ avoid any problems with disruption of nuclei and
  or distortions of the CMBR due to late decay or annihilations of TeV
  WIMP's.\footnote{$\mathcal{O}(10)$ GeV photons from $Q'$ annihilations are
  absorbed via $e^+ e^-$ production on the $T \sim  \mathcal{O}(30)$ eV gamma background
  existing at the time when $T' \sim$ 10 eV and lower energy $\sim$ GeV $\gamma$'s can
  thermalize by scattering on the CMBR. The exchange of gluons ``Sommerfeld" enhances
early perturbative annihilation by $\sim$ 3-10 which is, however, insufficient.}

  Unfortunately, KLN have not found an efficient universal mechanism that relaxes, also for
  neutral $X$'s, the extended $X\bar{X}$ bound states initially formed with large angular
  momenta $L$ to  $L \sim$ 0 where annihilations can happen.  (The two-photon relaxation
  discussed at length is rather inefficient and may require times of $\mathcal{O}$(sec).
  Furthermore, the estimate of \cite{Kang:2006yd} of the residual $Q'$ density after hadronically
  assisted ``late" annihilation (HAA) $Y(Q') \sim 10^{-18}$ is mistaken in the use they make
  of their Eq. (24):
\begin{equation}
  Y_X = \frac{n_{X}}{s} \sim 10^{-18}\times\left(\frac{R}{{\rm GeV}^{-1}}\right)^{-2}\times
  \left(\frac{T_B}{180 {\rm MeV}}\right)^{-3/2}\times \left(\frac{m_X}{{\rm TeV}}\right)^{1/2}.
\label{KLTeq24}  % {Eq x} Eq (1)
\end{equation}

  Specifically they use for $T_B$ the effective temperature when the $Q'\bar{Q'}$ states form,
  $T_B \sim $ 180 MeV $\sim$ the QCD confinement phase transition temperature.
  We find, however, a lower $T_B \sim$ 4.5 MeV yielding a 300 times larger residual $Y_X$.

  The point is that the pions and the muons and neutrinos into which they decay are, at such
  temperatures, in chemical equilibrium.  The initial huge number densities $(n_\pi \sim
  T^3)$, are comparable to those of photons {\it cannot} be neglected. These pions outnumber the
  $Q$'s by $10^{14}$ and the dissociation:
\begin{equation}
  \pi + Q'\bar{Q'} \rightarrow \bar{q}Q' + q\bar{Q'}
\label{pionsoutnumberQs} %Eq 2
\end{equation}
drives back the process of forming the $Q'\bar{Q'}$ states. The latter states form only
once the temperature  dropped enough so that the Boltzman factor
$\exp(-m_\pi/T)$ is $\sim 10^{-14}$ or $T \sim m_\pi$/30 $\sim$ 4.5 MeV.

  However a more careful study suggests yet another QCD  mechanism which accelerates the
  annihilations. Pions not only break up $Q'\bar{Q'}$ states but also induce an extremely fast
  relaxation to lower more tightly bound states, a key element missing in the KLN analysis.

  Specifically pion scattering can lead to ``diffractive de-excitation". While scattering
  from the excited and extended $Q'\bar{Q'}$ system via two gluon exchange, the pion can gain
  energy $\Delta(E)$ as the $Q'\bar{Q'}$ system cascades down to a more tightly bound state.
The cross section for diffractive de-excitation is smaller than that of the breakup of the
  initial state,yet it does qualitatively change the picture and largely resurrects the
  $\sim 10^{-3}-10^{-4}$ HAA suppression of $n_X=n_{Q'}$.
  The point is that only pions of energies larger than the binding of the  $ Q'\bar{Q'}$
  state can break it up, {\it any} pion can de-excite it. Further, as the bound state becomes
  more tightly bound there is even a greater chance that it will de-excite again before
  meeting a sufficiently high energy pion and dissociate.

    Once the Boltzman suppression $e^{B.E./T}$ compensates for the $\sim 10^{-2}-10^{-3}$ ratio
  of the cross sections for de-excitation and for breakup, the former becomes dominant. This
  suggests then that the effective temperature ``$T_B$" at which $Q'\bar{Q'}$ bound states
  form,  is around 40 MeV, smaller than $T_B$ = 180 MeV of KLN but $\sim$ 10 times larger
  than the small value of 4.5 MeV used above. When substituted in Eq. (\ref{KLTeq24}) it
  leads to a HAA suppression of $Q$'s by $10^{-3}$ which is sufficient for our
  purposes.

  Another mistake of KLN is that a $Q'\bar{u}$ is $Y_{\rm protons}/Y_{Q'} \sim 10^4$ more
  likely to meet first a proton and form $Q'ud + \bar{u}d$.
  Only later when $Q'ud$ meets a $\bar{Q'}u$ that a $Q'\bar{Q'}$ + proton will form.
  While all these reactions are exothermic, the last one is less so than the (unfortunately
  rather rare) $\bar{Q'}u + Q'\bar{d} \rightarrow  Q'\bar{Q'}$ + pion envisioned by KLN. The
  moderate ensuing slowdown  of the capture rate is likely to be offset by the fact
  that their assumed hadronic cross section $\sim$ 1 mb implicit in the choice of $R \sim$
  1/GeV in Eq. (\ref{KLTeq24}) is an underestimate by $\sim$ 10. In passing we note that
  for the special case
 of $Q$'s, the $g$'s which are readily emitted as massless gluons also facilitate a very efficient
 independent cascading down mechanism for the $Q\bar{Q'}$ bound states.

  \section{Potential Difficulties with Gluons and Glueballs of $SU(N')$} \label{sec:gluons}

  If $\Lambda' < $ MeV, the $N'^2$-1 new unconfined $g'$ gluons act when the CMBR temperature
  is $\sim T \sim$ MeV at times of $t \sim$ sec as new relativistic degrees of freedom. For
  this not to adversely affect the successful predictions of BBN \cite{BBN}, we need that
  their effect will not exceed that of one extra neutrino. We showed \cite{JacobyNussinov}
  that the dilution of $g$'s relative to neutrinos and photons makes $N'=2$ and even $N'=3$
  consistent with the BBN bounds.
  Consideration of the eventual formation of $SU(N')$ singlets bound by $g'$ strings exclude
  $N'=2$. Indeed, in this case as many $\bar{Q'}\bar{Q'}$ and $Q'Q'$ (anti-) baryons form at
  $T' \sim \Lambda'$ as $Q'\bar{Q'}$ mesons. The ``(anti-) baryons" carrying $\bar{3}$ and 3
  color representation then manifest as fractionally charged  heavy isotopes which as
  emphasized all along are strongly excluded by direct measurements. This fixes the allowed
  $N'$ to be 3.

As the $g'$ gas cools below $\Lambda'$,  $SU(3')$ confinement sets in and $g'g'$ color$'$
  singlet glueballs form. Quenched lattice QCD calculations, which are fully justified here,
  suggest several stable glueballs: $(0^{++}, \; 0^{(++)'}, \; 2^{++}, \; 0^{+-}, \; \mathcal
  {O}^{(+-)'}$ and a $1^{--}g'g'g'$ state with masses of order several $\Lambda'$. These
  could decay into two photons via two consecutive $Q'$ and ordinary quark $q$ loops
  connected by two QCD gluons. The decay rate
\begin{equation}
   \sim \frac{1}{(2\pi)^6 \cdot(m_{Q'} \cdot m_q)^{8}}\cdot m_{\rm gb'}^{17} \alpha'^2
  \alpha_{\rm em}^2 \alpha_{QCD}^4
\end{equation}
is minuscule, making the $gb$'s effectively stable. This generates a large number
density comparable to the photons in the MCBR of $gb$'s all with masses $\sim \Lambda' >$ 10 eV.
  In general this yields over-critical density, and at the lower limit a mix of hot dark
  matter which exceeds the WMAP bounds.

  However, it has been noted some time ago \cite{CHM}, that such DM of initially large density
  $\sim T'^3 \sim \Lambda'^3$ and no chemical potential, i.e., no conserved number of particles,
  efficiently cannibalizes itself  via the 3 $\rightarrow$ 2 process:
\begin{equation}
     gb'+ gb'+ gb' \rightarrow gb' + gb'
\label{3to2process}%Eq 3
\end{equation}
for the lowest $0^{++} \; gb$'s which carry no quantum numbers.  For static thermal
  equilibrium the reverse 2 $\rightarrow$ 3 process proceeds with the same rate and the
  fixed $T'$ leads to a fixed number density $n(gb') \sim T'^3$. However, the Hubble
  expansion keeps cooling the $gb$'s and the back reaction becomes less and less feasible.
  The ``cross section" for three-particle collisions is an unfamiliar concept. Yet we can use
  dimensional arguments to deduce that the above 3 $\rightarrow$ 2 dilutions keeps on going
  for many cycles. To find the freeze-out  $gb'$ density we ask: When does the rate of the
  $gb'$ decimating $g'^3 \rightarrow gb'^2$ processes,
\begin{equation}
    R = \frac{1}{n'}\frac{dn'}{dt} \sim -n'^2 \cdot \frac{v}{\Lambda'^{5}}
\label{decimatingprocess} %Eq 4
\end{equation}
become equal to the Hubble expansion rate? In the $gb'$ dominated era,
of interest the latter is: $H=(n' \cdot \Lambda')^{1/2}$.

  In the above
\begin{equation}
    v \sim \left( \frac{T'}{m_{\rm gb'}} \right)^{1/2} \sim \left( \frac {T'}{\Lambda'}
     \right)^{1/2}
\end{equation}
is the relative velocity and $(\Lambda')^{-5}$ provides the length$^5$ scale relevant to
  this hadronic three $gb'$ collision. For $n'$ we use an equilibrium density
\begin{equation}
    \sim T'^3 e^{-m_{\rm gb'}/T'}
\end{equation}
and $m_{ \rm gb'} \sim \Lambda'$.

   $R=H$ yields a limiting ``freeze-out" temperature $T'_f$ or $x =T'_f/\Lambda'$ given by
\begin{equation}
   e^{-3/2x} \cdot x^5 = \frac{\Lambda'}{M_{\rm Planck}} \sim 10^{-27} - 10^{-21}
\label{freezeoutTf} %Eq 5
\end{equation}
For $\Lambda'$ = 10 eV -10 MeV we find  $x \sim$ 1/(30) - 1/(20), respectively. The final
\begin{equation}
     Y_{\rm gb'} \sim \frac{n_{\rm gb'}}{s}  \sim e^{-1/x}
\end{equation}
yields negligible residual $gb'$ numbers and densities.

\section{The Residual $Q$'s After $SU(3')$ Confinement (Below $T' \sim \Lambda'$)}

  Invoking the late $T' \sim \Lambda' \; SU(3')$ confinement transition we show that (a) most of
  the $Q'$ and $\bar{Q'}$ relics annihilate, and (b) the remaining $Q'$ or $\bar{Q}'$'s are
  segregated into harmless charge and color neutral $(\bar{Q'})^3$ and $Q'^3$ Quirk  (anti-)
  baryons. We further verify that the latter $Q'^3$ ``baryons" do not bind to nuclei to form
  anomalous heavy isotopes.

    Rather than repeat here the detailed discussion \cite{JacobyNussinov},  we recall the chain of
arguments leading to the above conclusions dwelling on the more delicate steps.
In particular, we show that ordinary Coulombic repulsion is overcome in all
relevant cases by the $g'$ attractive exchange so that the final
annihilation of the $Q'$ and $\bar{Q'}$
connected by the $SU'(3)$ string cannot be evaded.

\begin{enumerate}
\item  The $SU(3')$ phase transition occurs at $T' \sim \Lambda'$.  We denote by $T',T$ the
  temperatures of the $g'-Q'$ sector and of the ordinary photons so that dilution effects make
  $T' \sim T/2$).  After $T'$ drops below $\Lambda'$ the $Q'$s and the $\bar{Q'}$s become
  connected by color$'$  flux tubes of tension $\sigma' \sim (\Lambda')^2$.
  Similar strings form between $Q'-\bar{Q'}$ pairs produced at LHC and in ultra high energy
  cosmic ray collisions.

  The hadronic (QCD) strings readily break by producing  $\bar{q}-q$
  in the strong field in the chromoelectric flux tube $E \sim \Lambda^2$. The rates contain
  the tunneling factor:
\begin{equation}
   \exp (-\pi \, m_q^2/(gE)) \sim  \exp (-[m_q/\Lambda]^2).
\label{tunnelfactor} %Eq 6
\end{equation}

The analog exponent for massive ($\mathcal{O}$(TeV)) $Q'$s and smaller $\Lambda'$ is huge
  and the strings connecting $Q'-\bar{Q'}$ pairs are unbreakable.  In vacuum the system then
  oscillates back and forth for a very long time until a rare close encounter leads to
  annihilation.

\item Initially the strings formed at $T \sim T' \sim \Lambda'$ between $Q'-\bar{Q'}$
  remnants are very long:
\begin{equation}
\ell_{init}(t; T'=\Lambda') \sim \frac{1}{\Lambda'} \cdot Y_{Q'}^{-1/3} \sim \frac{10^6}{\Lambda'}
\label{longstrings} %Eq 7
\end{equation}
as compared with an equilibrium distance between the confined $Q'-\bar{Q'}$ of
$\sim T'/\Lambda'^2 \sim 1/\Lambda'$. Thus we have to make sure that efficient relaxation
mechanisms (specifically the ``friction" due to Quirks colliding with the abundant $g'g'$
glueballs initially present) will pull the $Q'-\bar{Q'}$ back together to distances
$\ell_{final} \sim 1/\Lambda'$.  Our analysis using collision cross section  $1/\Lambda'^2$
and number and energy densities $\sim \Lambda'^3$ and $\Lambda'^4$ for the $gb$'s yielded
the relaxation time
\begin{equation}
   t_{relax} \sim \frac{M_{Q'}^{1/2}}{\Lambda'^{3/2}}.
\label{relaxtime} %Eq 8
\end{equation}

The key element is that this time  is shorter than the corresponding Hubble times:
$t_{Hub} \sim M_{Planck}/\Lambda'^2$ required for cooling from $T \sim \Lambda' \; T
\sim \Lambda'/2$, justifying  $T \sim \Lambda'$ used in $t_{Hub}$.
After such relaxation times the distance between the $\bar{Q'}$ and $Q'$ decreases to the
 normal expectation for thermal equilibrium  of $T'/\sigma' \sim  1/\Lambda'$.

\item In the above we referred to the color$'$ triplet sources to which the $SU(3')$
  strings are attached as $Q'-\bar{Q'}$. However, the remnants are not just $Q'$s or
  $\bar{Q'}$s but the QCD color singlet composites: $Q'ud$ and $\bar{Q'}u$. This does not
  affect the stages when the distance between those is $\ell > 1/\Lambda' >> 1/\Lambda$ where
  these ``Hadronic" composites can be viewed as point particles.

  For efficient decimation of the $Q'$ remnants the pull of
  the $SU(3')$ string should lead to the energetically favored rearrangement:
\begin{equation}
   Q'ud + \bar{Q'} u \rightarrow (Q'\bar{Q'}) + {\rm proton} \; (=uud)
\label{rearrangement} %Eq 9
\end{equation}
The $Q'\bar{Q'}$ state of a size smaller than  $\Lambda^{-1}$ then very quickly relaxes
  via the effectively massless perturbative $g'$ emission on time scales of order $10^{-17}$
  sec to the S wave ground state and annihilates.
  All we need then is to verify that the above rearrangement happens fast enough and does
  not constitute a dangerous bottleneck.  The $g'$ emission and attending
  shrinkage of the size of the system start before the light quarks have been ``etched out" and
  the heavy Quirkonium formed. All we need is that
  $\ell$ be of order $1/\Lambda'$ or smaller so that the $g'$ force transforms from the
  fixed tension to an $1/\ell^2$ increasing Coulombic force.  Even if
  conservatively the last effect is neglected we found a rearrangement rate:
  $\sigma_H \cdot \Lambda'^{7/2}/(M_{Q'}^{1/2})$ yielding rearrangement times of
$2 \cdot 10^5 - 2 \cdot 10^{-16}$ sec for $\Lambda' = 10 {\rm eV} - 10 {\rm MeV}$,
  where $\sigma_H$ is the ordinary hadronic cross sections $\mathcal{O}$(10) mbarn.
  Again these rearrangement times are far shorter than the respective Hubble times.

  The above arguments and those in the previous section \ref{sec:gluons} then ensure that
  {\it all} the (initially very extended) ``$Q'$"-$``\bar{Q'}$" states eventually annihilate.
  (We denote by ``$Q'$" the compound $SU(3)_c$ singlet $Q'ud$ and likewise
  $``\bar{Q'}$" =$\bar{Q'}u.$)

\item  At the time of the confinement transition of the color$'$ group (at temperatures
  $T' \sim \Lambda'$, $``Q'"^3 \; SU(3)$ baryonic singlets form in a few percent of the cases.
  We verified that the analog to stage 3 above rearrangements
\begin{equation}
   (Q'ud)^3 \rightarrow Q'^3 + {\rm proton} + {\rm neutron}
\end{equation}
are also completed at times much shorter than the Hubble time.
Unlike the $Q'\bar{Q'}$ mesons, isolated lowest Quirk baryons $Q'^3$ are stable.

    Because of the Quirk baryons and anti-baryons initially formed are extended, we expect
substantial annihilations, i.e., rearrangements of the two $Y$ shaped color$'$ networks
 of the baryon and anti-baryon into three $Q'\bar{Q'}$ mesons reducing the $Q'$ baryon
density far below $(n_{Q'}^3)/s  \sim  10^{-20}$.

\item A key observation is that  the few remnant $Q'^3$ baryons (and anti-baryons)
  pose no problem. The colored $\bar{Q'}u$ and $Q'ud$ remnants readily bind to nuclei
  forming  fractionally charged ultra heavy isotopes for which we have strong direct
  terrestrial bounds. However, the $Q'^3$ baryons are both color singlets and charge neutral.
  While Fermi statistics requires one $L=1$ $qq$ pair in the $Q'^3$ ground state, its size
  $\sim 1/(m_{Q'}\alpha')$ is very small. This in turn  makes for minuscule
  residual attractive Casimir Polder-like ordinary color forces and the $Q'^3$ baryons do
  not bind even to the heaviest/biggest nuclei and the dangerous ultra-heavy isotopes will
  not form. Also the non-dissipative Quirk baryons do not concentrate in the galactic
  discs/stars or planets. (Note that none of this in not true if $N' =2$
as then the $``Q'"^2$ baryons are colored and fractionally charged!)

\item A very important point not previously discussed in sufficient
  detail is the effect of ordinary Coulombic repulsion on $Q'-\bar{Q'}$ annihilations in
  the period when $T \sim T' < \Lambda'$.  We no longer have a $g'$ plasma
  and the new color$'$ forces are unscreened. When the separation between the ``$Q'$" and
  $``\bar{Q'}$" color singlet $Q'ud$ or $\bar{Q'}u$ composite is small $(\ell < 1/\Lambda'$),
  the $g'$ exchange interactions are Coulombic, $\sim (3/2) \cdot (\alpha'/\ell)$.
  The annihilation could potentially be blocked only if the
ordinary Coulomb repulsion $Z_1Z_2\alpha_{\rm em}/\ell$ overcomes this $g'$ exchange interaction
with $Z_1,Z_2$, the electric charges of the hadronic systems to which the $Q'$ and
  $\bar{Q'}$ are attached. (At large distances $\ell < 1/\Lambda'$ the constant color$'$
attraction clearly wins over the decreasing Coulomb blocking.)
The running of the $SU(3')$ color interaction is controlled by the $b_0$ term in the
  $\beta$ function which for the $n_f=0$ (no light Quirks) relevant here is $-11/(2\pi)
  \sim 1.8$. The smallest distance relevant for comparing $\alpha'$ and $\alpha_{\rm em}$
is $\sim$ 2 Fermi since for shorter distances ordinary nuclear forces dominate.
  For $\Lambda'$ = 10 eV-100 KeV this   distance is $10^{-7} - 10^{-3}/\Lambda'$
  and $\alpha'(\ell)$ ranges
between 0.08 to $\sim$ 0.04, namely $\alpha' \sim 11-5.5 \, \alpha_{\rm em}$. Thus
  if $ZZ' < 16-8$ we have $3/2 \alpha' > \alpha_{\rm em}, \; |F(g')| > F(em)$ and no
  blocking of $Q'-\bar{Q'}$ annihilations. The relic $Q'ud \; (Z=1/3)$ and $\bar{Q'}u
  \; Z=2/3$ clearly satisfy $ZZ' < 8-16$, and readily annihilate.

Big bang nucleosynthesis occurs at times of order 1-100 sec and temperature $\mathcal{O}$(1-0.1 MeV).

  In addition to Helium, also Lithium, Beryllium and Boron isotopes form though with
  tiny $\mathcal{O}$(10$^{-10})$ abundances relative to Hydrogen. Because of the
  Coulombic repulsion the positively charged $Q'ud$ and $\bar{Q'}u$ relics are
  unlikely to attach to such nuclei.\footnote{The $Q'ud \bar{Q'}u$  at BBN have
  temperatures/energies $\sim$ 0.1-1 MeV---very different from the energetic
  ($\mathcal{O}$ TeV!) $Q'ud$ and $\bar{Q'}-u$ or $\bar{Q'}-d$ emerging from the
  intersection points at the LHC and which may bind to the heavier nuclei encountered
  therein.}

  Even if some $Q'$ or $\bar{Q'}$-Li or Be composites  formed at the time of BBN, their
  mutual strong e.m.~repulsion suppresses binding pairs of those via a color$'$  string.
  The point is that at temperatures $T \sim T' > \Lambda' \sim$ 10 eV - 100 KeV we still
  have a gluon$'$ plasma but no longer the $e^+-e^-$ plasma. The gluon$'$ exchange is
  screened but {\it not} the ordinary e.m.~repulsion. Thus when the $SU(3')$ strings form,
  pairs of heavier nuclei with $Q'$ and $\bar{Q'}$ tend to be further separated decreasing
  the probability that these will be connected by strings connecting them.
  Rather, the more heavily charged $Q'$-Nucleus remnant will most likely bind
to a lighter $Z'<Z$ one.

This is over and above the fact that  the abundance of ``$Q'$" Be-$\bar{``Q'"}$Li and similar
heavy $Q'$ remnant pairs is anyway extremely small,
  $\sim  10^{-29}$ that of protons (recall $n_{Q'}/(n_p) \sim  10^{-9}$
  and $n_{{\rm Be,Li}}/n_p \sim 10^{-10}$) and  the $10^{10}$ more frequent
$``Q'"{\rm H}-\bar{``Q'"}$ Be, say, readily annihilate. Hence we conclude that
the annihilation is not hindered by Coulombic effects and the Quirk scenario
  weathers out this hurdle as well.
\end{enumerate}

\section{Cosmic Ray Produced Quirks}

Very high energy cosmic ray can produce in atmospheric collisions pairs of
stable TeV particles.  A negatively charged particle such as $X^-$ binds to nuclei and manifests
as an anomalously heavy isotope. The fact that very stringent $\sim  10^{-29}$ bounds on the abundance of
  TeV {\it Hydrogen} isotopes in ocean water accumulating therein over many
millions, even billion years occur,\cite{BKRref2and3} has been used by \cite{BKR} to limit
  such scenarios.   Since $Q'$ pairs are readily produced by UHE cosmic rays and bind
  to Hydrogen and Oxygen in water, one may wonder if the above bounds can directly
  exclude {\it all} Quirk scenarios irrespective of any cosmology?!

Of all the many issues discussed, this question (posed to us by W.~Marciano) may well
  be the single most relevant constraint on the model. Still, as we show next, the Quirks
  survive it as well.
  The saving grace is the confining $SU(3')$ string connecting the two nuclei to which the $Q'$
  and $\bar{Q'}$ are attached. We have various types of such connected pairs;  H--H, O--H, O--O,
  H--X, O--X, and X--X, with X some heavier nuclei. For simplicity we omitted the $Q'ud$ and
  $\bar{Q'}u$ at each vertex, and X denotes a generic solute nucleus such as Na, Cl, etc. Of
  all these combinations only the H--H is close to mimicking an ultra-heavy Hydrogen.

The initial total charge of the two Hydrogens and the $ud$ and $u$ attached to the $Q'$ and
  $\bar{Q'}$, respectively, is 3. The electrostatic interactions is likely to induce a
  $p \rightarrow n \; \beta$ transition so as to leave us with a system mimicking
  ultra-heavy Helium. The charges of  all combinations are $> \mathcal{O}(10)$ and
  misidentification as Hydrogen isotopes is unlikely. However, the low charges
  $Z=4/3 \; Z'=5/3$ in the H--H combination, the $g'$ exchange attraction
  discussed in the previous section overcomes Coulombic repulsion and
annihilation will quickly proceed. Thus the potentially dangerous and most restrictive
  upper bound on ultra heavy Hydrogen isotopes does not exclude the Quirk model
  and future ultra-sensitive limits on other isotopes of larger charge may
  eventually lead to exclusion  or discovery/model.

\section{A Short Comment on LHC Quirk Signatures}

  We studied in our previous paper the highly speculative possibility that the very long
  $> \mathcal{O}$ (100 Meter) strings, connecting $Q'-\bar{Q'}$ pairs to be hopefully
  produced at LHC can eventually be discovered. Such a discovery would have truly fantastic
  ramifications. It requires the lowest $SU(3')$ scales $\Lambda' = \mathcal{O}$(10 eV).
  On the other hand,  $Q'-\bar{Q'}$ and connecting string manifest in the LHC detectors
  for a wider range of 100 ev $< \Lambda' <$ 100 MeV.

As seen above, the $Q'-\bar{Q'}$ confinement, ensuing relaxation to short strings and eventual
  annihilation, dramatically accelerate as $\Lambda'$ increases. Thus the values pertinent for
  LHC signatures lead to much ``safer" cosmologically string scenarios as well.

As pointed out by Kang and Luty, a variety of LHC Quirk signatures at LHC arise for
  different $\Lambda'$. Let $T$, the $Q'-\bar{Q'}$ center of mass frame kinetic energy,
  be: $T \sim 1/m_{Q'} \sim $ TeV.  The maximal (often largely transverse) separation between
  the $Q'$ and $\bar{Q'}$:
\begin{equation}
     \ell \sim  T/\sigma' \sim  T/\Lambda'^2
\label{maxsep} %Eq 10
\end{equation}
varies for 100 eV $< \Lambda' <$ 100 MeV from $\sim$ 1 cm to 1 $\sim$ 10 Fermi. While
  completing at speed 0.7c its ``Yo-Yo" shaped orbit, the  $``Q'"-\bar{``Q'"}$ system,
  moves a distance $\sim \ell$, for each crossing. We would like to be more precise
  about the probability for and the nature of collisions of the $ud$ diquark and
  $u$ quark attached to the two end $Q'$ and $\bar{Q'}$.
  The systems here are the (very!) heavy quark (HQ) analogs of $\Lambda = sud$ and
  $K^+ = \bar{su}$. In the HQ limit these spin singlet states are almost degenerate
  with their hyperfine partners---the analogs of the $1^-K^{+*}$ and $3/2^+
  \Lambda$(1385).  The analogs to the substantial $K^* K \pi$ and
  $\Lambda(1385)\Lambda\pi$ couplings then imply an $\mathcal{O}(1)$ probability
  of producing one pion in each hadronic collision. This pion is almost at rest in
  the $``Q'" -\bar{``Q'"}$ system and carries
  away a kinetic energy  $\delta (T) \sim$ 150-300 MeV. Pion productions will then
  keep going on so long as the relative velocity of the heavy systems is substantial,
  say, $> c/2$, by which time $\sim$ 1/4 TeV, namely half of the initial kinetic energy
  has been ``radiated" away via $\mathcal{O}(100)$ pions.
  Since a $uu$ or $dd$ diquark is energetically disfavored, a $\pi^+(or \pi^-)$ is
  likely to emerge from  the $\bar{Q'}u$ end with a $u \rightarrow d$ transition
  or from $\bar{Q'}d$ with a $d \rightarrow u$ transition.
  If a pion is produced in each crossing we expect a remarkable signature of alternating
  sign pions correlated with alternate (upward or downward) motion between spatially
  (and hence temporally) consecutive emissions.

The cross section for pion production is $\sim (1/\Lambda_{\rm QCD})^2$ whereas
  the physical cross section of the $SU(3')$ flux tube is $\sim 1/\Lambda'^2$.
  The probability $P$ for producing a pion in a given encounter is therefore:
\begin{equation}
     P=\left(\frac{\Lambda'}{\Lambda}\right)^2
\end{equation}
and $1/P$ crossings are required on average for one pion production.  Combining with
  the previous equation for $\ell$, the distance between consecutive
  crossings, we  find that consecutive pion production events are separated by
\begin{equation}
    \Delta L \sim \ell/P \sim T \frac{\Lambda^2}{\Lambda'^4} \sim 10 \; {\rm Fermi-cm}
\label{DeltaL} %Eq 11
\end{equation}
for $\Lambda'$ ranging between 100 MeV to 100 KeV.

\end{document}